\documentclass[aps,prl,reprint,notitlepage,superscriptaddress,nofootinbib,floatfix]{revtex4-2}

\usepackage{amsmath}
\usepackage{amssymb}
\usepackage{bm}
\usepackage{mathtools}
\usepackage{tikz}
\usetikzlibrary{arrows.meta,calc,positioning,patterns}
\usepackage{hyperref}
\hypersetup{colorlinks=true,allcolors=blue}

\newcommand{\ii}{\mathrm{i}}
\newcommand{\ketvac}{\lvert0\rangle}
\definecolor{quartetteal}{RGB}{0,112,118}
\definecolor{pairgold}{RGB}{181,116,0}
\definecolor{electronred}{RGB}{170,52,58}
\definecolor{rkblue}{RGB}{54,91,145}

\begin{document}

\title{Direct Charge-4e Superconductivity by Mobilizing Bound Quartets}
\author{Jin-Tao Jin}
\affiliation{Department of Physics, The Hong Kong University of Science and Technology, Clear Water Bay, Kowloon 999077, Hong Kong, China}
\author{Pengfei Li}
\affiliation{Department of Physics, The Hong Kong University of Science and Technology, Clear Water Bay, Kowloon 999077, Hong Kong, China}
\author{Yi Zhou}
\affiliation{Institute of Physics, Chinese Academy of Sciences, Beijing 100190, China}
\date{\today}

\begin{abstract}
A direct charge-\(4e\) superconductor exhibits coherent four-electron order without condensation in any charge-\(2e\) pairing channel. We establish three complementary results. First, a particle-number-conserving Hamiltonian with electron hopping and two-body interactions moves bound quartets efficiently: for electron- and pair-hopping amplitudes \(t\) and \(K\) and local gap \(U_0\), virtual pair transfer produces quartet hopping at order \(K^2/U_0\), whereas electron hopping alone contributes only at order \(t^4/U_0^3\). Second, the microscopic theory identifies an exactly solvable parent Hamiltonian at the phase-separation boundary. Its ground state combines quartet off-diagonal long-range order and finite phase stiffness without pair condensation, while rigorous bounds establish quartet order beyond the parent in a family with explicit quartet transfer. Third, a single convergent strong-coupling expansion simultaneously controls quartet kinetics and thermodynamic stability and rules out condensation in every pair channel of the microscopic model. On bipartite hypercubic lattices at half quartet filling, it yields a parametrically controlled homogeneous phase for \(d\geq2\), with quartet order, a linear collective mode, and gapped charge-\(2e\) excitations. Virtual pair transfer thus enables a direct charge-\(4e\) superconducting ground state without pair condensation.
\end{abstract}

\maketitle

Whether coherence can emerge first in a composite channel while all lower-charge channels remain uncondensed is a fundamental question in interacting quantum matter. Direct charge-\(4e\) superconductivity gives this question a sharp electronic form: bound quartets acquire phase rigidity, yet no charge-\(2e\) pairing channel condenses. A microscopic realization requires mobile quartets and a thermodynamically homogeneous phase in a particle-number-conserving Hamiltonian. These requirements can compete: strong binding can suppress quartet motion, whereas the virtual processes that mobilize quartets can also generate density interactions favoring density order or phase separation.

Kivelson, Emery, and Lin proposed charge-\(4e\) superconductivity in the weak-hopping regime of the two-dimensional \(t\)-\(J\)-\(V\) model, where \(t\), \(J\), and \(V\) denote hopping, exchange, and density interaction. They identified mobile charge-\(4e\) plaquette molecules competing with charge-\(2e\) superfluidity and phase separation; their companion study found that the pure \(t\)-\(J\) model favors electron pairs over larger clusters \cite{KivelsonEmeryLin1990PRB,EmeryKivelsonLin1990PRL}. Later theories obtain charge-\(4e\) order as a vestige of pair-density-wave, multicomponent, or nematic pairing \cite{Agterberg2008,Berg2009,Fradkin2015,Agterberg2020,Wu2024,Huecker2026,Fernandes2021,Jian2021,Zou2026}. Larkin--Ovchinnikov states can melt through dislocation proliferation into quartet superfluids without pair condensation \cite{RadzihovskyVishwanath2009,Radzihovsky2011}; intercomponent drag can stabilize charge-\(4e\) order without charge-\(2e\) condensation \cite{Herland2010}; and vestigial order from spin-triplet pairing in superfluid \({}^{3}\mathrm{He}\) offers a neutral-superfluid analogue \cite{Volovik2024}. Direct quartet phases also occur in interacting lattice models \cite{Soldini2024,Wan2026,Gao2026,Shi2026,Jiang2017}, while a solvable many-flavor construction realizes charge-\(4e\) order with explicit particle-number breaking \cite{Gnezdilov2022}.

Here we establish an analytically controlled route from bound quartets to a homogeneous direct charge-\(4e\) superconductor satisfying three requirements: quartet mobility, thermodynamic stability, and no pair condensation. Pair hopping overcomes the kinetic suppression of electron-only quartet transfer, while a separate density interaction tunes stability. An exact parent has a quartet-condensed ground state with finite stiffness but lies at the phase-separation boundary, anchoring a nearby strong-coupling regime with gapped, uncondensed pairs. For the microscopic Hamiltonian, containing only electron hopping and two-body interactions, a single convergent expansion controls quartet kinetics, density curvature, and the full two-particle density matrix. At zero temperature in \(d\geq2\), the largest eigenvalues of the full four- and two-particle density matrices provide particle-number-conserving diagnoses of quartet and pair condensation \cite{Yang1962}; quartet stiffness measures phase rigidity, and inverse quartet compressibility measures thermodynamic stability.

\textbf{Microscopic model.---}
We consider a periodic \(d\)-dimensional hypercubic lattice of \(N_s\) cells with even linear size and coordination \(z_{\mathrm c}=2d\). In cell \(\bm R\), \(c_{\bm R,m}\) annihilates an electron in mode \(m=1,\cdots,4\). Define \(\hat n_{\bm R,m}=c_{\bm R,m}^{\dagger}c_{\bm R,m}\), \(\hat n_{\bm R}=\sum_m\hat n_{\bm R,m}\), and the pair creation operators \(p_{\bm R,+}^{\dagger}=c_{\bm R,1}^{\dagger}c_{\bm R,2}^{\dagger}\) and \(p_{\bm R,-}^{\dagger}=c_{\bm R,3}^{\dagger}c_{\bm R,4}^{\dagger}\). Let \(U_0>0\) set the local gap scale, \(K\) and \(t\) be real pair- and electron-hopping amplitudes, and \(V_{\mathrm{nn}}\) be the nearest-neighbor density interaction. The onsite term is
\[
 \begin{aligned}
 H_0&=\frac{U_0}{4}\sum_{\bm R}\hat n_{\bm R}(4-\hat n_{\bm R})\\
 &=\frac{3U_0}{4}\sum_{\bm R}\hat n_{\bm R}
 -\frac{U_0}{2}\sum_{\bm R,m<n}\hat n_{\bm R,m}\hat n_{\bm R,n}.
 \end{aligned}
\]
This form contains only one- and two-body onsite terms. Summing over nearest-neighbor bonds \(\langle\bm R\bm R'\rangle\), the full Hamiltonian is
\begin{align}
 H={}&H_0+V_{\mathrm{nn}}\sum_{\langle\bm R\bm R'\rangle}
 (\hat n_{\bm R}-2)(\hat n_{\bm R'}-2)\nonumber\\
 &-K\sum_{\langle\bm R\bm R'\rangle,\eta=\pm}
 (p_{\bm R,\eta}^{\dagger}p_{\bm R',\eta}+\mathrm{H.c.})\nonumber\\
 &-t\sum_{\langle\bm R\bm R'\rangle,m}
 (c_{\bm R,m}^{\dagger}c_{\bm R',m}+\mathrm{H.c.}),
 \label{eq:microscopic}
\end{align}
where \(\eta=\pm\) labels the two pair channels and H.c.\ denotes the Hermitian conjugate. Thus the full microscopic Hamiltonian contains only one-body electron hopping and two-body interactions.

A cell with \(n=0,\cdots,4\) electrons has onsite energy \(E_n=U_0n(4-n)/4\). The empty and quartet states form a zero-energy doublet connected by \(q_{\bm R}^{\dagger}=p_{\bm R,+}^{\dagger}p_{\bm R,-}^{\dagger}\); the one- and three-electron states cost \(3U_0/4\), while the six two-electron states form a virtual sector at energy \(U_0\). We use \(\lambda\ll1\) with \(|K|,|t|=O(\lambda U_0)\) and \(V_{\mathrm{nn}}=O(\lambda^2U_0)\), imposing the sharper condition \(t^2\ll|K|U_0\).

Projecting onto the zero-energy doublet, define \(n_{\bm R}^{(q)}=q_{\bm R}^{\dagger}q_{\bm R}\) and \(\bm T_{\bm R}=(T_{\bm R}^{x},T_{\bm R}^{y},T_{\bm R}^{z})\) by \(T_{\bm R}^{+}=q_{\bm R}^{\dagger}\), \(T_{\bm R}^{-}=q_{\bm R}\), and \(T_{\bm R}^{z}=n_{\bm R}^{(q)}-1/2\), where \(T_{\bm R}^{\pm}=T_{\bm R}^{x}\pm\ii T_{\bm R}^{y}\) and \(\ii\) is the imaginary unit. Writing \(E_{\mathrm c}\) for an additive constant and \(\delta H_{>2}\) for local terms beyond \(O(\lambda^2)\), degenerate perturbation theory gives the effective XXZ Hamiltonian
\begin{align}
 H_{\mathrm{eff}}={}&E_{\mathrm c}
 +J_z\sum_{\langle\bm R\bm R'\rangle}
 T_{\bm R}^{z}T_{\bm R'}^{z}\nonumber\\
 &-J_\perp\sum_{\langle\bm R\bm R'\rangle}
 (T_{\bm R}^{x}T_{\bm R'}^{x}
 +T_{\bm R}^{y}T_{\bm R'}^{y})
 +\delta H_{>2},\label{eq:effective-xxz}\\
 J_\perp={}&\frac{2K^2}{U_0},\qquad
 J_z=16V_{\mathrm{nn}}+\frac{2K^2}{U_0}+\frac{16t^2}{3U_0}.
 \label{eq:couplings}
\end{align}
A fixed sector with \(M\) quartets has \(N_e=4M\) electrons and filling \(\nu_4=M/N_s\). At zero chemical potential, particle--hole symmetry forbids a quartet-density field and identifies half filling as the symmetry point; we work at this filling in the fixed sector \(\nu_4=1/2\).

\textbf{Quartet mobility and stability are distinct.---}
Within the empty/quartet doublet, the XXZ terms describe hard-core quartet bosons with hopping \(t_b=J_\perp/2\) and nearest-neighbor interaction \(V_b=J_z\). Denoting pair- and electron-hopping contributions by superscripts \((K)\) and \((t)\), pair hopping transfers a quartet in two virtual steps, giving \(t_b^{(K)}=K^2/U_0\) and \(V_b^{(K)}=2K^2/U_0\). Electron hopping contributes to quartet transfer only at fourth order, \(t_b^{(t)}=16t^4/(3U_0^3)\), but generates a density interaction at second order, \(V_b^{(t)}=16t^2/(3U_0)\). Pair hopping therefore removes the kinetic disadvantage of electron-only motion: quartet kinetics and the induced interaction share the scale \(K^2/U_0\), while the intermediate charge-\(2e\) states remain gapped at \(O(U_0)\).

On a bipartite graph, a sublattice rotation maps Eq.~\eqref{eq:effective-xxz} to the antiferromagnetic XXZ convention with anisotropy \(\Delta=J_z/J_\perp\). At half quartet filling, the leading nearest-neighbor model is easy-plane for \(\lvert J_z\rvert<J_\perp\), or \(-1<\Delta<1\). With only pair hopping (\(V_{\mathrm{nn}}=t=0\)), the model lies at the upper boundary \(\Delta=1\); attractive \(V_{\mathrm{nn}}\) accesses the sufficient leading-order interval
\begin{equation}
 -\frac{K^2}{4U_0}-\frac{t^2}{3U_0}
 <V_{\mathrm{nn}}<-\frac{t^2}{3U_0}.
 \label{eq:easy-plane-window}
\end{equation}
This window has width \(K^2/(4U_0)\), set by pair hopping; \(V_{\mathrm{nn}}\) tunes the system through it, while electron hopping shifts both boundaries toward stronger attraction by \(t^2/(3U_0)\). Higher-order and longer-range terms determine the finite-coupling boundary, and at dilute or incommensurate fillings the easy-plane regime may extend beyond this half-filled interval.

\textbf{Exact parent and quartet condensate.---}
At leading order, the lower edge of Eq.~\eqref{eq:easy-plane-window}, \(V_{\mathrm{nn}}^*=-K^2/(4U_0)-t^2/(3U_0)\), is exactly solvable. Let \(P_{\bm R}^{(0)}\) and \(P_{\bm R}^{(4)}\) project onto empty and full cells, define \(P_{\bm R}=P_{\bm R}^{(0)}+P_{\bm R}^{(4)}\), and extend \(\bm T_{\bm R}\) by zero to partially occupied cells. For a bond state \(|n,n'\rangle\) with \(n\) and \(n'\) electrons on its two cells, define \(|a\rangle=(|4,0\rangle-|0,4\rangle)/\sqrt2\). With positive onsite and bond scales \(U\) and \(J\), the exact parent Hamiltonian is
\begin{equation}
 \begin{aligned}
 H_{\mathrm{exact}}&=U\sum_{\bm R}(1-P_{\bm R})
 +J\sum_{\langle\bm R\bm R'\rangle}h_{\bm R\bm R'},\\
 h_{\bm R\bm R'}&=\frac14P_{\bm R}P_{\bm R'}
 -\bm T_{\bm R}\cdot\bm T_{\bm R'}
 =|a\rangle\langle a|.
 \end{aligned}
 \label{eq:exact-parent}
\end{equation}
This particle-number-conserving parent is finite range but not few body: its bond projector contains an explicit eight-fermion quartet-transfer term. At \(V_{\mathrm{nn}}^*\), Eq.~\eqref{eq:couplings} gives \(J_z=-J_\perp\) and \(J=J_\perp=2K^2/U_0\), so the leading projected microscopic Hamiltonian matches \(H_{\mathrm{exact}}\) bond by bond.

Let \(\ketvac\) denote the fermion vacuum. On any connected graph, \(H_{\mathrm{exact}}\) has a unique zero-energy state in each fixed sector \(N_e=4M\), namely the uniform quartet Dicke state
\begin{equation}
 |\Psi_M\rangle=
 \frac{1}{M!\sqrt{\binom{N_s}{M}}}
 \left(\sum_{\bm R}q_{\bm R}^{\dagger}\right)^M\ketvac .
 \label{eq:dicke-state}
\end{equation}
The \(N=4\) \(\eta\)-clustering state and a gauge-equivalent parent on bipartite lattices were introduced by Yoshida and Katsura \cite{Yoshida2022}, within the broader \(\eta\)-pairing lineage \cite{Yang1989,deBoer1995,Montorsi1996}. Here we generalize the minimal two-term uniform parent to any connected graph, providing an exact boundary for the microscopic mechanism.

Let \(\lambda_{\max}^{(k)}\) denote the largest eigenvalue of the full \(k\)-particle density matrix in the basis of normalized antisymmetric fermion products. For \(|\Psi_M\rangle\),
\begin{equation}
 \lambda_{\max}^{(4)}=\frac{M(N_s-M+1)}{N_s},\qquad
 \lambda_{\max}^{(2)}=\frac{M}{N_s}.
 \label{eq:quartet-density-spectrum}
\end{equation}
At fixed \(0<\nu_4<1\), \(\lambda_{\max}^{(4)}=O(N_s)\) establishes quartet condensation through off-diagonal long-range order (ODLRO), while \(\lambda_{\max}^{(2)}=O(1)\) excludes pair condensation. On a periodic nearest-neighbor hypercubic lattice of linear size \(L\), let \(\rho_4(L,M)\) denote the quartet phase stiffness obtained from a boundary twist. Exactly,
\begin{equation}
 \rho_4(L,M)=J\frac{M(N_s-M)}{N_s(N_s-1)}
 \longrightarrow J\nu_4(1-\nu_4).
 \label{eq:exact-quartet-stiffness}
\end{equation}
Its nonzero thermodynamic limit establishes quartet superfluidity in addition to condensation. The paramagnetic current annihilates the equal-amplitude Dicke state configuration by configuration, leaving only the diamagnetic contribution (see End Matter). Writing \(E_0(N_e)\) for the ground-state energy at fixed electron number, \(E_0(4M)=0\) for every \(M\), so the inverse quartet compressibility vanishes despite the finite stiffness. Thus the exact parent is a quartet superfluid but remains thermodynamically marginal. The following section establishes the fixed-sector dynamical exponent \(z=2\).

\begin{figure}[!tb]
 \centering
 \begin{tikzpicture}[
   x=1cm,y=1cm,
   panel/.style={draw=black!22,fill=black!0.6,rounded corners=1.2pt,line width=0.35pt},
   cell/.style={draw=black!55,fill=white,rounded corners=0.7pt,minimum width=4.8mm,minimum height=4.8mm,inner sep=0pt},
   full/.style={cell,fill=quartetteal!88,draw=quartetteal!90!black}
 ]
  \begin{scope}[xscale=1.85,yscale=1.00]
   \draw[panel] (-0.18,-0.15) rectangle (4.33,5.28);
   \node[anchor=north west,font=\bfseries\small]
     at (0.05,5.14) {RK and exclusion-process structure};
   \node[font=\footnotesize] at (2.08,4.39)
     {four-cell open path at \(M=2\)};
   \node[font=\footnotesize,text=black!65] at (2.08,4.04)
     {edges: one nearest-neighbor quartet hop};

   \node[inner sep=1pt] (c1100) at (2.08,3.65)
     {\tikz[baseline=-0.5ex,x=4.8mm,scale=0.86,transform shape]{\node[full] at (0,0) {};\node[full] at (1,0) {};\node[cell] at (2,0) {};\node[cell] at (3,0) {};}};
   \node[inner sep=1pt] (c1010) at (2.08,3.10)
     {\tikz[baseline=-0.5ex,x=4.8mm,scale=0.86,transform shape]{\node[full] at (0,0) {};\node[cell] at (1,0) {};\node[full] at (2,0) {};\node[cell] at (3,0) {};}};
   \node[inner sep=1pt] (c1001) at (1.10,2.55)
     {\tikz[baseline=-0.5ex,x=4.8mm,scale=0.86,transform shape]{\node[full] at (0,0) {};\node[cell] at (1,0) {};\node[cell] at (2,0) {};\node[full] at (3,0) {};}};
   \node[inner sep=1pt] (c0110) at (3.06,2.55)
     {\tikz[baseline=-0.5ex,x=4.8mm,scale=0.86,transform shape]{\node[cell] at (0,0) {};\node[full] at (1,0) {};\node[full] at (2,0) {};\node[cell] at (3,0) {};}};
   \node[inner sep=1pt] (c0101) at (2.08,2.00)
     {\tikz[baseline=-0.5ex,x=4.8mm,scale=0.86,transform shape]{\node[cell] at (0,0) {};\node[full] at (1,0) {};\node[cell] at (2,0) {};\node[full] at (3,0) {};}};
   \node[inner sep=1pt] (c0011) at (2.08,1.45)
     {\tikz[baseline=-0.5ex,x=4.8mm,scale=0.86,transform shape]{\node[cell] at (0,0) {};\node[cell] at (1,0) {};\node[full] at (2,0) {};\node[full] at (3,0) {};}};

   \draw[rkblue,line width=0.7pt] (c1100) -- (c1010);
   \draw[rkblue,line width=0.7pt] (c1010) -- (c1001);
   \draw[rkblue,line width=0.7pt] (c1010) -- (c0110);
   \draw[rkblue,line width=0.7pt] (c1001) -- (c0101);
   \draw[rkblue,line width=0.7pt] (c0110) -- (c0101);
   \draw[rkblue,line width=0.7pt] (c0101) -- (c0011);

   \draw[rkblue!55!black,line width=0.45pt]
     (0.28,1.22) -- (3.88,1.22);
   \node[align=center,font=\footnotesize,text=rkblue!80!black]
     at (2.08,0.80)
     {quartet-hop graph \(\Rightarrow\) exclusion-process Laplacian\\
      \(\Rightarrow\) equal amplitudes \((|\Psi_M\rangle)\)};
   \node[align=center,font=\footnotesize] at (2.08,0.18)
     {spatial graph
      \(\Rightarrow\) exact fixed-sector gap
      \(\sim J L^{-2}\),\quad \(z=2\)};
  \end{scope}
 \end{tikzpicture}
 \caption{\label{fig:exclusion-dynamics} Exact exclusion-process dynamics
 in the fixed-\(M\) empty/quartet sector, illustrated for \(M=2\) on a
 four-cell open path. Filled (open) squares denote quartet-occupied
 (empty) cells, and edges connect configurations related by one
 nearest-neighbor quartet hop. The equal-amplitude superposition is the
 zero mode \(\lvert\Psi_M\rangle\). On a finite-range periodic lattice,
 the spatial graph Laplacian fixes the exact gap in this sector, which
 scales as \(L^{-2}\) (\(z=2\)).}
\end{figure}

\textbf{Exact fixed-sector dynamics and gap.---}
In each fixed-\(M\) empty/quartet manifold, the projected parent has Rokhsar--Kivelson (RK) stochastic-matrix form \cite{Rokhsar1988,Henley2004,Castelnovo2005,Ardonne2004}. Viewing a quartet-occupied cell as a particle and an empty cell as a vacancy, each bond term contributes an off-diagonal matrix element \(-J/2\) for their exchange and a matching diagonal escape rate \(J/2\). Every fixed-\(M\) block is therefore exactly the symmetric simple exclusion process (SSEP) Markov Laplacian \(\mathcal L_M\), whose conventional master-equation generator is \(-\mathcal L_M\) [Fig.~\ref{fig:exclusion-dynamics}]. Connectivity of the configuration graph makes its constant zero mode unique \cite{Seneta1981}. Let \(G\) denote the interaction graph, \(L_G\) its spatial graph Laplacian, and \(\lambda_2(L_G)\) the first positive eigenvalue of \(L_G\). For \(1\leq M\leq N_s-1\), let \(\Delta_M^{(0,4)}\) and \(\Delta_M\) denote the gaps at fixed \(N_e=4M\) within the empty/quartet manifold and the full Hilbert space, respectively. Then
\begin{equation}
 \Delta_M^{(0,4)}=\frac J2\lambda_2(L_G),\quad
 \Delta_M=\min\left\{\frac J2\lambda_2(L_G),2U\right\}.
 \label{eq:ssep-gap}
\end{equation}
The first equality, independent of \(M\), is the SSEP corollary of Aldous' spectral-gap theorem, proved by Caputo, Liggett, and Richthammer \cite{Caputo2010}. The scale \(2U\) is the exact energy of the lightest charge-compatible states with partially occupied cells, which contain two such cells. On a finite-range periodic lattice of linear size \(L\), \(\lambda_2(L_G)\sim L^{-2}\), so the full gap scales as \(L^{-2}\) at fixed \(0<\nu_4<1\), establishing \(z=2\); this gaplessness is consistent with filling constraints \cite{Lieb1961,Oshikawa2000,Hastings2004}. The finite quartet phase stiffness and vanishing inverse quartet compressibility place the parent at the \(\Delta=-1\) isotropic-ferromagnet point, the boundary of phase separation.

\textbf{Rigorous quartet order beyond the parent.---}
Deforming the exact boundary toward the easy-plane side yields a finite-range fermionic family with explicit quartet transfer [End Matter, Eq.~\eqref{eq:exact-easy-plane-family}]. For a sufficiently large onsite penalty, rigorous spin-model bounds establish quartet ODLRO at the hypercubic XY point for \(d\geq2\) and, at fixed half quartet filling, throughout \(0\leq\Delta<0.20\) on the square lattice \cite{Kennedy1988,Kubo1988,Ozeki1989}; in both cases, \(\lambda_{\max}^{(2)}\leq1\) excludes pair condensation. This independent nonperturbative benchmark complements the lower-body microscopic route below, whose quartet order is established separately.

\begin{figure}[!tb]
 \centering
 \begin{tikzpicture}[
   x=0.98cm,y=1cm,
   region/.style={rounded corners=1.4pt,line width=0.55pt},
   phase/.style={align=center,font=\footnotesize},
   minor/.style={align=center,font=\footnotesize},
   axis/.style={-{Latex[length=1.8mm,width=1.2mm]},black!65,line width=0.55pt},
   flow/.style={-{Latex[length=1.6mm,width=1.1mm]},line width=0.75pt}
 ]
  \node[anchor=west,font=\bfseries\small] at (0.10,8.15)
   {(a) Leading nearest-neighbor XXZ cut};

  \filldraw[region,draw=electronred!78!black,fill=electronred!12]
   (0.10,6.10) rectangle (2.25,7.30);
  \filldraw[region,draw=quartetteal!82!black,fill=quartetteal!12]
   (2.25,6.10) rectangle (5.95,7.30);
  \filldraw[region,draw=pairgold!82!black,fill=pairgold!13]
   (5.95,6.10) rectangle (8.25,7.30);
  \node[phase,text=electronred!82!black] at (1.18,6.70)
   {\textbf{phase}\\[1pt]\textbf{separation}};
  \node[phase,text=quartetteal!82!black] at (4.10,6.73)
   {\textbf{easy-plane quartet}\\[1pt]\textbf{regime}, \(z=1\)};
  \node[phase,text=pairgold!82!black] at (7.10,6.73)
   {\textbf{quartet}\\[1pt]\textbf{density order}};

  \draw[rkblue!90!black,line width=1.1pt] (2.25,5.98) -- (2.25,7.42);
  \filldraw[fill=rkblue,draw=white,line width=0.5pt]
   (2.25,6.70) circle (2.6pt);
  \node[anchor=south east,font=\footnotesize,text=rkblue!90!black]
   at (2.25,7.39) {\textbf{exact parent}};
  \node[anchor=south west,font=\footnotesize,text=rkblue!90!black]
   at (2.25,7.39) {:\ \(\Delta=-1,\ z=2\); first-order boundary};
  \draw[axis] (0.10,5.78) -- (8.34,5.78);
  \node[anchor=north,font=\scriptsize] at (8.34,5.69) {\(\Delta\)};
  \draw[black!65,line width=0.55pt] (2.25,5.69) -- (2.25,5.87);
  \draw[black!65,line width=0.55pt] (5.95,5.69) -- (5.95,5.87);
  \node[minor] at (1.18,5.51) {\(\Delta<-1\)};
  \node[minor] at (2.25,5.51) {\(-1\)};
  \node[minor] at (4.10,5.51) {\(-1<\Delta<1\)};
  \node[minor] at (5.95,5.51) {\(1\)};
  \node[minor] at (7.10,5.51) {\(\Delta>1\)};

  \node[anchor=west,font=\bfseries\small] at (0.10,4.95)
   {(b) Microscopic neighborhood of the parent};
  \filldraw[region,draw=electronred!78!black,fill=electronred!12]
   (0.10,2.15) rectangle (2.55,3.45);
  \filldraw[region,draw=black!35,fill=black!4]
   (2.55,2.15) rectangle (8.25,3.45);
  \filldraw[region,draw=black!55,fill=black!12,densely dotted]
   (2.275,2.15) rectangle (2.825,3.45);
  \filldraw[region,draw=rkblue!88!black,fill=rkblue!14,dashed]
   (2.825,2.15) rectangle (5.55,3.45);
  \node[phase,text=electronred!82!black] at (1.19,2.80)
   {\textbf{phase}\\[1pt]\textbf{separation}\\[1pt]\(\kappa_4^{-1}<0\)};
  \node[phase,text=rkblue!90!black] at (4.19,2.80)
   {\textbf{controlled}\\[1pt]\textbf{stability wedge}\\[1pt]
    \(z=1\) dynamics};
  \node[phase,text=black!65] at (6.90,2.80)
   {outside\\asymptotic\\control};

  \draw[rkblue!90!black,dash dot,line width=0.9pt]
   (2.55,2.03) -- (2.55,3.55);
  \draw[flow,electronred!88!black]
   (2.40,4.35) -- node[above,font=\scriptsize] {flow} (0.70,4.35);
  \draw[flow,quartetteal!88!black]
   (2.70,4.35) -- node[above,font=\scriptsize] {flow} (4.40,4.35);
  \node[minor,text=rkblue!90!black] at (2.55,3.78)
   {cubic XXZ parent tuning};
  \draw[axis] (0.10,1.98) -- (8.34,1.98);
  \node[anchor=north,font=\scriptsize] at (2.55,1.92) {\(0\)};
  \node[anchor=north east,font=\scriptsize] at (8.25,1.92)
   {\(V_{\mathrm{nn}}-V_{\mathrm{nn}}^{*,(3)}\)};

  \draw[black!55,densely dotted,line width=0.8pt]
   (0.45,1.34) -- (1.05,1.34);
  \node[anchor=west,font=\scriptsize] at (1.15,1.34)
   {boundary layer:
    \(\|\delta H_{\mathrm{rem}}\|_{\mathrm{loc}}\sim|\kappa_4^{-1}|\)};
  \draw[rkblue!88!black,dashed,line width=0.8pt]
   (0.45,0.96) -- (1.05,0.96);
  \node[anchor=west,font=\scriptsize,text=rkblue!90!black] at (1.15,0.96)
   {controlled:
    \(\|\delta H_{\mathrm{rem}}\|_{\mathrm{loc}}
    \ll\kappa_4^{-1}\ll J_\perp\)};
 \end{tikzpicture}
 \caption{\label{fig:xxz-stability} Leading XXZ structure and controlled
 microscopic stability at half quartet filling. (a) The nearest-neighbor
 model is phase separated for \(\Delta<-1\), easy plane for
 \(-1<\Delta<1\), and quartet-density ordered for \(\Delta>1\). The exact
 parent at \(\Delta=-1\) is a first-order boundary with \(z=2\) dynamics.
 (b) The dash-dotted line is the cubic-order parent tuning
 \(V_{\mathrm{nn}}^{*,(3)}\) [Eq.~\eqref{eq:shifted-parent}], not the full
 microscopic phase boundary. The dashed wedge obeys
 \(\|\delta H_{\mathrm{rem}}\|_{\mathrm{loc}}\ll
 \kappa_4^{-1}\ll J_\perp\); here \(\delta H_{\mathrm{rem}}\) is the
 remainder starting at fourth order,
 \(\|\cdot\|_{\mathrm{loc}}\) is the local interaction norm, and
 \(\kappa_4^{-1}>0\) is the density curvature. Arrows show
 renormalization-group flow away from this tuning. At zero
 temperature, the wedge supports
 \(z=1\) quartet order for \(d\geq2\) and a
 quartet Luttinger liquid for \(d=1\). The dotted strip marks the boundary
 layer where the remainder competes with the density curvature; the
 light-gray region is outside asymptotic control.}
\end{figure}

\textbf{A controlled homogeneous phase.---}
We now return to the microscopic Hamiltonian \(H\); Figure~\ref{fig:xxz-stability} contrasts the leading XXZ phase structure with the controlled microscopic neighborhood of the parent. In the Hermitian des Cloizeaux convention \cite{desCloizeaux1960}, every cubic process on a bipartite graph that returns to the empty/quartet manifold is bond local and only renormalizes \(J_\perp\) and \(J_z\). Absorbing these \(Kt^2/U_0^2\) corrections shifts the parent tuning to
\begin{equation}
 V_{\mathrm{nn}}^{*,(3)}
 =-\frac{K^2}{4U_0}-\frac{t^2}{3U_0}-\frac{8Kt^2}{9U_0^2}.
 \label{eq:shifted-parent}
\end{equation}
Henceforth, \(J_\perp\) and \(J_z\) denote these cubic-renormalized couplings. Through cubic order, the detuning from the parent is \(\delta_{\mathrm{ep}}\equiv J_z+J_\perp=16(V_{\mathrm{nn}}-V_{\mathrm{nn}}^{*,(3)})\). The remaining operators begin at fourth order; we collect them in \(\delta H_{\mathrm{rem}}\), whose local interaction norm the convergent linked-cluster expansion bounds uniformly in system size, \(\|\delta H_{\mathrm{rem}}\|_{\mathrm{loc}}=O(\lambda^4U_0)\) \cite{Bravyi2011}. Let \(e_0(\nu_4)\) be the ground-state energy per cell. Near the parent, the density curvature is
\begin{equation}
 \kappa_4^{-1}\equiv\frac{\partial^2e_0}{\partial\nu_4^2}
 =z_{\mathrm c}\delta_{\mathrm{ep}}+o(\delta_{\mathrm{ep}})
 +O(\lambda^4U_0).
 \label{eq:inverse-compressibility}
\end{equation}
Let \(c_n=O(J_\perp)\) be the parent's positive density-gradient coefficient and, measuring momentum in inverse lattice-spacing units, let \(k=|\bm k|\) and \(k_\times=(\kappa_4^{-1}/c_n)^{1/2}\). Three scales then control the density response: positive curvature for \(k\lesssim k_\times\), the \(c_nk^2\) term for \(k_\times\lesssim k\ll1\), and the exact parent scale \(O(J_\perp)\) for lattice momenta bounded away from zero. In the controlled regime of Eq.~\eqref{eq:controlled-stability-wedge} below, each exceeds the \(O(\lambda^4U_0)\) remainder, excluding both macroscopic phase separation and a finite-wave-vector density instability.

Let \(\tau\) denote imaginary time, \(\bm x\) position, \(\phi(\bm x,\tau)\) the quartet phase, \(n_4=\nu_4+\delta n_4\) the local quartet density, and \(\rho_4\) the quartet phase stiffness. For smooth, topologically trivial configurations, the background Berry term \(\ii\nu_4\partial_\tau\phi\) is a total derivative. With \(\hbar\) set to one, the long-wavelength action is
\begin{align}
 \mathcal S=\int d\tau\,d^dx\biggl[
 &\ii\delta n_4\partial_\tau\phi+
 \frac{\kappa_4^{-1}}2(\delta n_4)^2\nonumber\\
 &+\frac{\rho_4}2(\bm\nabla\phi)^2+
 \frac{c_n}2(\bm\nabla\delta n_4)^2+\cdots\biggr].
 \label{eq:phase-density-action}
\end{align}
At the parent, \(\rho_4=J_\perp\nu_4(1-\nu_4)\) and \(c_n=J_\perp/[4\nu_4(1-\nu_4)]\). The Gaussian equations of motion give the collective-mode dispersion
\begin{equation}
 \omega^2(\bm k)=\rho_4|\bm k|^2
 \left(\kappa_4^{-1}+c_n|\bm k|^2\right).
 \label{eq:hydrodynamic-dispersion}
\end{equation}
At the parent, where \(\kappa_4^{-1}=0\), \(\omega(\bm k)=J_\perp k^2/2\), reproducing the exact \(z=2\) branch. Equation~\eqref{eq:detuning-rg} shows that the dimensionless curvature detuning \(\bar\delta\) is a relevant perturbation of the \(z=2\) parent and grows in magnitude under coarse-graining. For positive curvature, below \(k_\times\), it produces a linear \(z=1\) mode with velocity \(v_4=(\rho_4\kappa_4^{-1})^{1/2}\), whereas negative curvature drives phase separation.

The controlled regime is
\begin{equation}
 \|\delta H_{\mathrm{rem}}\|_{\mathrm{loc}}
 \ll\kappa_4^{-1}\ll J_\perp\ll U_0.
 \label{eq:controlled-stability-wedge}
\end{equation}
Taking a representative positive density curvature, \(\kappa_4^{-1}=O(\lambda^3U_0)\), makes this hierarchy parametrically nonempty; up to fixed coordination factors, it corresponds microscopically to \(\lambda^4U_0\ll V_{\mathrm{nn}}-V_{\mathrm{nn}}^{*,(3)}\ll\lambda^2U_0\). The lower bound protects positive curvature against the remainder, while \(\kappa_4^{-1}\ll J_\perp\) separates continuum and lattice scales. Smooth local analytic nonlinearities are irrelevant at the \(z=1\) Gaussian theory. Compact defects, such as vortices and phase slips, are controlled separately by the isotropized stiffness \(\mathcal K_{\mathrm{iso}}\equiv\sqrt{\rho_4\kappa_4}\), where \(\kappa_4=(\kappa_4^{-1})^{-1}\) is the quartet compressibility. For the representative scaling, \(\mathcal K_{\mathrm{iso}}=O(\lambda^{-1/2})\), placing the compact \(z=1\) theory in its ordered large-stiffness regime for \(d\geq2\) and asymptotically suppressing defect fugacities. These conditions establish the asymptotic stability wedge in Fig.~\ref{fig:xxz-stability}(b) and quartet ODLRO within it for \(d\geq2\).

\textbf{Pair motion without pair condensation.---}
Electron and pair hopping break the exact parent's cellwise \(\mathbb Z_4\) selection rule, namely conservation of each cell's electron number modulo four; hence the absence of pair condensation no longer follows from symmetry. The convergent local Schrieffer--Wolff expansion instead supplies the required locality bounds (see End Matter) \cite{Bravyi2011}. Let \(E_0(N_e)\) now denote the ground-state energy of \(H\) with \(N_e\) electrons. On a regular bipartite graph of coordination \(z_{\mathrm c}\), the charge-\(2e\) excitation gap is
\begin{align}
 \Delta_{2e}(M)&=E_0(4M+2)-\frac{E_0(4M)+E_0(4M+4)}2\nonumber\\
 &=U_0-z_{\mathrm c}|K|+O(\lambda^2U_0)>0.
 \label{eq:pair-gap-expansion}
\end{align}
The remainder is uniform in system size and \(M\). Pair defects thus disperse with bandwidth \(O(K)\) yet retain an \(O(U_0)\) gap, while local dressing produces nonzero but exponentially decaying intercell pair correlations. A separate bound on the largest eigenvalue of the full two-particle density matrix gives
\begin{equation}
 \lambda_{\max}^{(2)}=O(1),
 \label{eq:microscopic-pair-density-bound}
\end{equation}
excluding condensation in every two-electron orbital, not only in the onsite pair channels.

Combining quartet order in the controlled wedge with Eqs.~\eqref{eq:pair-gap-expansion} and \eqref{eq:microscopic-pair-density-bound} establishes a direct charge-\(4e\) superconducting ground state at half quartet filling and zero temperature for \(d\geq2\): quartet ODLRO without pair condensation. For \(d=1\), the same regime is a quartet Luttinger liquid \cite{Luther1975,Haldane1980,Giamarchi2003}. The linear mode belongs to the ungauged matter theory; gauging the global \(U(1)\) converts the finite quartet phase stiffness \(\rho_4\) into a charge-\(4e\) London response. Outside the controlled wedge, finite-momentum density order or pair-gap closure can intervene at generic finite coupling.

\textbf{Conclusion and outlook.---}
These results yield a threefold design principle: bind quartets locally, mobilize them through virtual pair transfer, and tune their density interaction to the homogeneous side of the parent boundary. Complementary finite-coupling studies find direct quartet phases in one-dimensional models and in attractive \(SU(4)\) and bond-centered Su--Schrieffer--Heeger (SSH) models \cite{Soldini2024,Wan2026,Shi2026}. Because the SSH interaction generates pair transfer together with density, exchange, and correlated-hopping terms, relating it quantitatively to \(K\) and \(U_0\) requires a dedicated strong-coupling projection. Quantum Monte Carlo and tensor-network simulations of Eq.~\eqref{eq:microscopic} could directly test the predicted phase and map its stability window at finite coupling.

\begin{acknowledgments}
This work was supported in part by the National Key Research and Development Program of China (Grant No.~2022YFA1403403) and the National Natural Science Foundation of China (Grant Nos.~12274441 and 12534004). J.-T.~J. and P.F.L. are supported by a fellowship award from the Hong Kong Research Grants Council (Project No.~SRFS2324-6S01) and the New Cornerstone Science Foundation.
\end{acknowledgments}

\textit{Data availability.---}No external datasets were used. The Python and Wolfram Language scripts used for exact diagonalization and algebraic verification are available from the corresponding author upon reasonable request.

\bibliography{reference}

\clearpage
\section*{End Matter}

\textbf{Exclusion dynamics and exact gap.---}
Within the fixed-\(M\) empty/quartet manifold, each active bond projector
acts in the basis \(\{|4,0\rangle,|0,4\rangle\}\) as
\(\tfrac12\bigl(\begin{smallmatrix}1&-1\\-1&1\end{smallmatrix}\bigr)\).
The fixed-sector parent block is therefore the SSEP Markov Laplacian
\(\mathcal L_M\) with hop rate \(J/2\), distinct from the spatial graph
Laplacian \(L_G\). Connectivity of the configuration graph makes the
constant vector its unique zero mode. Because the SSEP is the
label-invariant restriction of the interchange process, Aldous'
spectral-gap theorem implies a gap of at least
\(J\lambda_2(L_G)/2\). Conversely, for a quartet configuration
\(\mathcal C\), a spatial eigenmode
\(L_G\phi_\alpha=\lambda_\alpha\phi_\alpha\) produces the SSEP
eigenfunction
\(F_\alpha(\mathcal C)=\sum_{\bm R\in\mathcal C}\phi_\alpha(\bm R)\).
Choosing \(\lambda_\alpha=\lambda_2(L_G)\) saturates the bound, so the
projected gap equals \(J\lambda_2(L_G)/2\), independently of \(M\)
\cite{Caputo2010}. The lightest fixed-charge states outside the
empty/quartet manifold contain two partially occupied cells and cost
\(2U\). Combining the two sectors gives Eq.~\eqref{eq:ssep-gap}; on
finite-range periodic lattices, \(\lambda_2(L_G)\sim L^{-2}\),
establishing the exact \(z=2\) branch.

\textbf{Exact quartet phase stiffness.---}
On a periodic \(L^d\) hypercubic lattice, introduce a total quartet
twist \(\Theta\) along \(\hat x\) through
\[
 \begin{aligned}
 H_x(\Theta)&=-\frac J2\sum_{\bm R}
 \left(e^{\ii\Theta/L}q_{\bm R+\hat x}^{\dagger}q_{\bm R}
 +\mathrm{H.c.}\right),\\
 \rho_4&=L^{2-d}E_0''(0).
 \end{aligned}
\]
Here \(E_0(\Theta)\) is the fixed-\(M\) ground-state energy. For
untwisted fixed-sector eigenstates \(\{|n\rangle\}\), with energies
\(E_n\) and \(|0\rangle=|\Psi_M\rangle\), nondegeneracy gives
\[
 E_0''(0)=\langle H_x''(0)\rangle
 -2\sum_{n\ne0}
 \frac{|\langle n|H_x'(0)|\Psi_M\rangle|^2}{E_n-E_0}.
\]
In every occupation configuration contributing to \(|\Psi_M\rangle\),
periodicity equates the numbers of \(1\to0\) and \(0\to1\) domain
walls along each \(x\) row. Hence
\(H_x'(0)|\Psi_M\rangle=0\) configuration by configuration, and the
paramagnetic term vanishes. On each of the \(N_s\) positively
oriented \(x\) bonds,
\(\langle q_{\bm R+\hat x}^{\dagger}q_{\bm R}\rangle
=M(N_s-M)/[N_s(N_s-1)]\); the remaining diamagnetic term therefore
gives Eq.~\eqref{eq:exact-quartet-stiffness}.

\textbf{Exact density curvature and boundary.---}
Detuning the leading XXZ model from the parent by
\(\delta_{\mathrm{ep}}\) adds
\(\delta_{\mathrm{ep}}\sum_{\langle\bm R\bm R'\rangle}
T_{\bm R}^zT_{\bm R'}^z\). For distinct cells, its expectation value
in the Dicke state is
\begin{equation}
 \langle T_{\bm R}^zT_{\bm R'}^z\rangle
 =\left(\nu_4-\frac12\right)^2
 -\frac{\nu_4(1-\nu_4)}{N_s-1}.
 \label{eq:dicke-density-curvature}
\end{equation}
For \(\delta_{\mathrm{ep}}>0\), the energy is minimized by the sector
or sectors nearest half quartet filling; for
\(\delta_{\mathrm{ep}}<0\), by the empty or fully occupied sector. For
even \(N_s\), the right and left derivatives of the ground-state
energy per bond at \(\delta_{\mathrm{ep}}=0\) are
\(-1/[4(N_s-1)]\to0\) and \(1/4\), respectively. Their discontinuity
establishes a first-order coexistence boundary in the thermodynamic
limit. The same expectation value fixes the linear coefficient in
Eq.~\eqref{eq:inverse-compressibility}.

\textbf{Rigorous order in the explicit-transfer family.---}
On a regular bipartite lattice, for \(U,J>0\) and XXZ anisotropy
\(\Delta\), define
\begin{align}
 h_{\bm R\bm R'}^{\mathrm{ep}}(\Delta)
 &=\frac14P_{\bm R}P_{\bm R'}
 -T_{\bm R}^xT_{\bm R'}^x-T_{\bm R}^yT_{\bm R'}^y
 +\Delta T_{\bm R}^zT_{\bm R'}^z,\nonumber\\
 H_{\mathrm{ep}}(\Delta)
 &=U\sum_{\bm R}(1-P_{\bm R})
 +J\sum_{\langle\bm R\bm R'\rangle}
 h_{\bm R\bm R'}^{\mathrm{ep}}(\Delta).
 \label{eq:exact-easy-plane-family}
\end{align}
At \(\Delta=-1\), \(H_{\mathrm{ep}}\) reduces to
\(H_{\mathrm{exact}}\); \(\Delta=0\) is the XY point. A partially
occupied cell is annihilated by \(P_{\bm R}\) and all
\(\bm T_{\bm R}\), thereby deleting its incident bond terms. Because
deleting one XY bond can lower the energy by at most \(3J/4\),
\(U>3z_{\mathrm c}J/4\) forces the ground state into the all-active
sector. On periodic hypercubic lattices with \(d\geq2\) and even
linear size, the projected XY ground state is unique and, by
spin-flip symmetry, lies at half quartet filling. After undoing the
sublattice rotation, the Kennedy--Lieb--Shastry theorem establishes
uniform quartet ODLRO
\cite{Kennedy1988}:
\begin{equation}
 \liminf_{N_s\to\infty}\frac1{N_s^2}
 \sum_{\bm R,\bm R'}
 \langle q_{\bm R}^{\dagger}q_{\bm R'}\rangle>0.
 \label{eq:rigorous-quartet-odlro}
\end{equation}
Choosing the normalized uniform quartet orbital as a trial vector
gives
\[
 \frac{\lambda_{\max}^{(4)}}{N_s}\geq
 \frac1{N_s^2}\sum_{\bm R,\bm R'}
 \langle q_{\bm R}^{\dagger}q_{\bm R'}\rangle.
\]
Equation~\eqref{eq:rigorous-quartet-odlro} therefore establishes an
extensive four-particle eigenvalue in Yang's sense \cite{Yang1962}.
By contrast, intercell pair transfer leaves the empty/quartet
manifold, while intracell orbital changes are Pauli blocked. The full
two-particle density matrix is therefore diagonal with
\(\lambda_{\max}^{(2)}\leq1\).

On the square lattice, rigorous XXZ bounds extend transverse order
over a finite anisotropy interval: Kubo--Kishi establish it for
\(0\leq\Delta<0.13\), and Ozeki--Nishimori--Tomita for
\(0\leq\Delta<0.20\) \cite{Kubo1988,Ozeki1989}. At fixed half quartet
filling, replacing the blocked cells by empty or quartet-occupied
cells while preserving total charge embeds each compatible state into
the all-active sector. Since the spectral norm (largest absolute
eigenvalue) of the Hermitian two-cell bond operator is
\(\lVert h_{\bm R\bm R'}^{\mathrm{ep}}(\Delta)\rVert=(3-\Delta)/4\),
the blocked-sector comparison excludes partially occupied cells when
\begin{equation}
 0\leq\Delta<0.20,\qquad
 U>\frac{z_{\mathrm c}J}{4}(3-\Delta).
 \label{eq:xxz-blocked-bound}
\end{equation}
Thus, for \(z_{\mathrm c}=4\), this explicit-transfer family has
\(\lambda_{\max}^{(4)}=O(N_s)\) and \(\lambda_{\max}^{(2)}\leq1\) at
half quartet filling. Its XY point coincides with the leading
microscopic projection at
\(V_{\mathrm{nn}}=-K^2/(8U_0)-t^2/(3U_0)\), but the microscopic wedge
also contains generated longer-range and multibody terms and is
established separately.
The fixed-sector quartet Dicke state also belongs to a degenerate
Richardson--Gaudin family \cite{Richardson1963,Gaudin1976,Dukelsky2004};
related quartet correlations occur in nuclear and \(\alpha\)-particle
systems \cite{Dukelsky2006,Lerma2007,Ropke1998,Tohsaki2001} and
four-component one-dimensional gases \cite{Wu2005,Lecheminant2005}.

\textbf{Microscopic projection and local remainder.---}
Let \(P^{(0,4)}=\prod_{\bm R}P_{\bm R}\). At second order, pair
hopping contributes \(-K^2/U_0\) to every matrix element in the
anti-aligned bond basis \(\{|4,0\rangle,|0,4\rangle\}\), whereas
electron hopping contributes only the diagonal elements
\(-8t^2/(3U_0)\). The aligned states receive no hopping correction at
this order. Adding the projected density interaction
\(16V_{\mathrm{nn}}\sum_{\langle\bm R\bm R'\rangle}
T_{\bm R}^zT_{\bm R'}^z\) yields the couplings in
Eq.~\eqref{eq:couplings}.

We adopt the Hermitian des Cloizeaux convention, obtained by
symmetrically orthonormalizing the projected exact low-energy states;
this fixes the convention-dependent higher-order couplings
\cite{desCloizeaux1960}. On a bipartite graph, the complete cubic
corrections are
\begin{equation}
 \delta J_\perp=\frac{80}{9}\frac{Kt^2}{U_0^2},\qquad
 \delta J_z=\frac{16}{3}\frac{Kt^2}{U_0^2}.
 \label{eq:kt2-correction}
\end{equation}
Cell parity confines every surviving \(Kt^2\) process to one bond,
whereas pure \(K^3\) and \(t^3\) processes require triangular loops and
therefore vanish. Cubic order thus only renormalizes the
nearest-neighbor XXZ couplings and shifts the parent tuning as in
Eq.~\eqref{eq:shifted-parent}. The remaining connected clusters begin
at fourth order and satisfy the system-size-independent bound
\(\|\delta H_{\mathrm{rem}}\|_{\mathrm{loc}}=O(\lambda^4U_0)\)
throughout the convergent linked-cluster expansion \cite{Bravyi2011}.
Within the hierarchy of Eq.~\eqref{eq:controlled-stability-wedge},
positive curvature controls \(k\lesssim k_\times\), the parent term
\(c_nk^2=O(J_\perp k^2)\) controls \(k_\times\lesssim k\ll1\), and the
positive parent scale \(O(J_\perp)\) controls momenta away from zero.
Each scale dominates the local remainder for sufficiently small
\(\lambda\), excluding a finite-wave-vector density instability
throughout the controlled wedge.

\textbf{Relevant flow and compact-defect control.---}
Under \(z=2\) scaling, let \(\ell\) denote the logarithmic RG scale and
define the dimensionless detuning
\(\bar\delta=\kappa_4^{-1}/(c_n\Lambda^2)\), where \(\Lambda\) is the
momentum cutoff. Its tree-level flow and associated crossover length
\(\xi_\times\) are
\begin{equation}
 \frac{d\bar\delta}{d\ell}=2\bar\delta+\cdots,
 \qquad
 \xi_\times\sim\left(\frac{c_n}{\kappa_4^{-1}}\right)^{1/2}.
 \label{eq:detuning-rg}
\end{equation}
Positive curvature drives the crossover to the \(z=1\) phase theory,
where charge \(U(1)\) permits only derivative interactions, which are
irrelevant at the Gaussian fixed point. Compact defects are controlled after rescaling
imaginary time by the isotropized stiffness
\begin{equation}
 \mathcal K_{\mathrm{iso}}=\sqrt{\rho_4\kappa_4}
 \sim\sqrt{\frac{J_\perp}{\kappa_4^{-1}}}.
 \label{eq:isotropized-stiffness}
\end{equation}
For \(\kappa_4^{-1}=O(\lambda^3U_0)\),
\(\mathcal K_{\mathrm{iso}}=O(\lambda^{-1/2})\), suppressing defect
fugacities and placing the compact theory in the ordered
large-stiffness regime for \(d\geq2\).

\textbf{Uniform pair-gap and correlation estimates.---}
The lowest \(4M+2\) manifold contains \(M\) quartets and one
two-electron defect with zeroth-order cost \(U_0\). Pair hopping
mobilizes two of its six local flavors on each
\(z_{\mathrm c}\)-regular bipartite component of the one-defect
configuration graph, whose band minimum is
\(-z_{\mathrm c}|K|\); the other four remain immobile at first order.
The adjacent \(4M\) and \(4M+4\) sectors have no \(O(K)\) energy shift.
Cancellation of the common bulk energy in the parity-gap combination,
together with locality of the remaining connected clusters, therefore
gives
\begin{equation}
 \left|\Delta_{2e}(M)-U_0+z_{\mathrm c}|K|\right|
 \leq C(z_{\mathrm c})\lambda^2U_0,
 \label{eq:pair-gap-remainder}
\end{equation}
uniformly in \(N_s\) and \(0\leq M<N_s\) throughout the convergence
domain \cite{Bravyi2011}.

A local pair
\(P_{\bm R,ab}^{\dagger}=c_{\bm R,a}^{\dagger}c_{\bm R,b}^{\dagger}\),
\(a<b\), obeys
\(P^{(0,4)}P_{\bm R,ab}^{\dagger}P^{(0,4)}=0\). Consequently, in the
Schrieffer--Wolff-dressed pair correlator, every nonzero contribution
must contain a connected cluster joining \(\bm R\) and \(\bm R'\).
Spanning their graph distance \(r=d_G(\bm R,\bm R')\) costs at least
one pair hop or two electron hops per step. Convergence of the cluster
sum therefore gives, with correlation length \(\xi_2\) and
system-size-independent \(C_0\),
\begin{align}
 |\langle P_{\bm R,ab}^{\dagger}P_{\bm R',cd}\rangle|
 &\leq C_0e^{-r/\xi_2},\nonumber\\
 e^{-1/\xi_2}
 &=O\left(\frac{z_{\mathrm c}|K|}{U_0}
 +\frac{z_{\mathrm c}t^2}{U_0^2}\right)<1,
 \label{eq:pair-correlation-bound}
\end{align}

For an unordered pair \(X=\{i,j\}\) of distinct single-particle
orbitals, with a fixed canonical ordering, define
\(B_X^\dagger=c_i^\dagger c_j^\dagger\) and the full two-particle density
matrix \(\Gamma^{(2)}_{X,Y}=\langle B_X^\dagger B_Y\rangle\). On the
bounded-degree pair-orbital graph generated by elementary pair and
electron hops, let \(d_2(X,Y)\) be the graph distance. The same
expansion gives, for \(X\neq Y\), with correlation length
\(\widetilde\xi_2\) and system-size-independent \(C_2\),
\begin{align}
 |\Gamma^{(2)}_{X,Y}|
 &\leq C_2e^{-d_2(X,Y)/\widetilde\xi_2},\nonumber\\
 e^{-1/\widetilde\xi_2}
 &=O\left[\frac{z_{\mathrm c}(|K|+|t|)}{U_0}\right].
 \label{eq:pair-density-kernel-bound}
\end{align}
The linear \(|t|\) arises because one electron hop is an edge of this
graph, whereas Eq.~\eqref{eq:pair-correlation-bound} requires two.
The pair-orbital graph has bounded degree, so for sufficiently small
\(\lambda\) the exponential decay dominates the growth of its distance
shells and bounds every absolute row sum independently of \(N_s\).
Together with
\(\Gamma^{(2)}_{X,X}=\langle n_i n_j\rangle\leq1\), this bounds the
largest eigenvalue uniformly in system size, yielding
Eq.~\eqref{eq:microscopic-pair-density-bound}. Thus virtual admixtures
produce short-range pair correlations but no charge-\(2e\) ODLRO.

\clearpage
\widetext
\begin{center}
 \textbf{Supplemental Material}\\[3pt]
 \textbf{Direct Charge-\(4e\) Superconducting Ground State from Bound Quartets}
\end{center}
\twocolumngrid
\setcounter{equation}{0}
\renewcommand{\theequation}{S\arabic{equation}}
\renewcommand{\theHequation}{supp.\arabic{equation}}

\section*{S1. Exact boundary: order, spectrum, and phase stiffness}
\label{app:density-matrices}

\textbf{Exact density-matrix spectra.---}
The local projectors are
\begin{equation}
 P_{\bm R}^{(0)}=\prod_m(1-\hat n_{\bm R,m}),\qquad
 P_{\bm R}^{(4)}=\prod_m\hat n_{\bm R,m}.
 \label{eq:local-projectors}
\end{equation}
The normalized configuration states
\(\lvert\mathcal C\rangle=\prod_{\bm R\in\mathcal C}q_{\bm R}^{\dagger}\ketvac\)
give \(|\Psi_M\rangle=\binom{N_s}{M}^{-1/2}\sum_{|\mathcal C|=M}|\mathcal C\rangle\).
For \(1\leq a<b\leq4\), define the local pair operator
\(P_{\bm R,ab}^{\dagger}=c_{\bm R,a}^{\dagger}c_{\bm R,b}^{\dagger}\).
For distinct cells,
\begin{align}
 \langle q_{\bm R}^{\dagger}q_{\bm R'}\rangle
 &=\frac{M(N_s-M)}{N_s(N_s-1)},\label{eq:quartet-odlro}\\
 \langle P_{\bm R,ab}^{\dagger}P_{\bm R',cd}\rangle&=0,\label{eq:pair-zero}\\
 \langle n_{\bm R}^{(q)}n_{\bm R'}^{(q)}\rangle_{\mathrm{conn}}
 &=-\frac{\nu_4(1-\nu_4)}{N_s-1}.
 \label{eq:density-corr}
\end{align}
The exact parent is invariant under independent cellwise transformations
\(c_{\bm R,m}\mapsto\ii c_{\bm R,m}\). This cellwise \(\mathbb Z_4\)
symmetry isolates the local quartet block and enforces
Eq.~\eqref{eq:pair-zero}. The quartet block has diagonal \(M/N_s\) and
uniform off-diagonal value from Eq.~\eqref{eq:quartet-odlro}. Its uniform
eigenvalue is \(M(N_s-M+1)/N_s\); the \(N_s-1\) transverse eigenvalues
are \(M(M-1)/[N_s(N_s-1)]\). All other four-particle sectors are
diagonal in normalized antisymmetric products, with eigenvalues at most
one. The uniform eigenvalue is therefore the largest eigenvalue for
every nonempty sector; at complete filling it equals one and can be
degenerate.

The full two-particle matrix is diagonal in the normalized antisymmetric
pair basis: cellwise support forbids off-diagonal elements that transfer
charge between cells, while Pauli blocking removes off-diagonal elements
within a full cell. Its same-cell eigenvalues are \(M/N_s\), and its
distinct-cell eigenvalues are \(M(M-1)/[N_s(N_s-1)]\). This establishes
Eq.~\eqref{eq:quartet-density-spectrum}. Electron and pair hopping remove
the artificial cellwise rule; a generic charge-\(4e\) ordered state
retains only the global residual \(\mathbb Z_4\) associated with broken
charge \(U(1)\).

\textbf{Ground space and pseudospin symmetry.---}
\label{app:ground-space}
Every term in \(H_{\mathrm{exact}}\) is positive semidefinite, so a zero
mode lies in the kernel of each term. The onsite projectors first
restrict every cell to the empty/full manifold. There
\(h_{\bm R\bm R'}=(1-\mathsf P_{\bm R\bm R'})/2\), where \(\mathsf P\)
exchanges two pseudospins. Edge transpositions of a connected graph
generate all permutations. Hence a fixed-\(M\) zero mode has equal
amplitudes and is uniquely \(|\Psi_M\rangle\). Across all charge sectors,
\begin{equation}
 \mathcal G=\operatorname{span}\{|\Psi_M\rangle:0\leq M\leq N_s\},
 \qquad \dim\mathcal G=N_s+1.
 \label{eq:ground-space}
\end{equation}
The products \(\prod_{\bm R}(u|0\rangle_{\bm R}+v|4\rangle_{\bm R})\), \(|u|^2+|v|^2=1\), form a spanning subset of this space, not its full set of vectors.

Define the global quartet-pseudospin generators
\(S^+=\sum_{\bm R}q_{\bm R}^{\dagger}\),
\(S^-=(S^+)^\dagger\), and
\(S^z=\sum_{\bm R}T_{\bm R}^z\). The \(N_s+1\) Dicke zero modes form
one maximal-pseudospin \(SU(2)\) multiplet with \(S=N_s/2\). Their
cross-sector degeneracy is therefore symmetry generated rather than
topological.

\textbf{Exclusion dynamics and exact spectrum.---}
\label{app:parent-spectrum}
Within the fixed-\(M\) manifold,
\begin{align}
 h_{\bm R\bm R'}=\frac12\big[
 &n_{\bm R}^{(q)}(1-n_{\bm R'}^{(q)})
 +(1-n_{\bm R}^{(q)})n_{\bm R'}^{(q)}\nonumber\\
 &-q_{\bm R}^{\dagger}q_{\bm R'}
 -q_{\bm R'}^{\dagger}q_{\bm R}\big].
 \label{eq:smf-form}
\end{align}
This is stochastic-matrix form \cite{Henley2004,Castelnovo2005,Ardonne2004}. The fixed-\(M\) parent block equals the symmetric simple exclusion process (SSEP) Markov Laplacian \(\mathcal L_M\): each allowed quartet hop has rate \(J/2\), off-diagonal entries are minus the rates, diagonal entries sum outgoing rates, and columns sum to zero. For the vector \(p(\tau)\) of configuration probabilities at stochastic time \(\tau\), the master equation is \(\partial_\tau p=-\mathcal L_Mp\). Configuration-graph connectivity and Perron--Frobenius positivity independently give the unique constant zero mode \cite{Seneta1981}.

For the spatial graph \(G\), the one-particle Laplacian \(L_G=D_G-A_G\), with degree matrix \(D_G\) and adjacency matrix \(A_G\), is distinct from \(\mathcal L_M\). Caputo--Liggett--Richthammer prove equality of the interchange-process and random-walk gaps \cite{Caputo2010}. The SSEP is the restriction of the interchange process to label-invariant functions, so its gap is no smaller than the interchange gap. Conversely, \(L_G\phi_\alpha=\lambda_\alpha\phi_\alpha\) implies that \(F_\alpha(\mathcal C)=\sum_{\bm R\in\mathcal C}\phi_\alpha(\bm R)\) is an SSEP eigenfunction with eigenvalue \(J\lambda_\alpha/2\), furnishing the matching upper bound and proving the first equality in Eq.~\eqref{eq:ssep-gap}.

Call a cell blocked when its occupation is one, two, or three, and let
\(\mathcal B\) denote the blocked set. Quartet dynamics acts on the
active induced graph \(G\setminus\mathcal B\), and the corresponding
invariant block is
\begin{equation}
 H_{\mathrm{exact}}\big|_{\mathcal B}
 =U|\mathcal B|
 +J\sum_{\substack{\langle\bm R\bm R'\rangle\\
 \bm R,\bm R'\notin\mathcal B}}h_{\bm R\bm R'}.
 \label{eq:blocked-sector-decomposition}
\end{equation}
If the active graph is disconnected, the quartet number is conserved
separately on each component, and componentwise Dicke states attain the
lower edge \(U|\mathcal B|\). At total charge \(4M\), the smallest
compatible blocked set contains two cells, with occupations \((1,3)\)
or \((2,2)\), and therefore costs \(2U\). This proves the second
relation in Eq.~\eqref{eq:ssep-gap} when combined with the projected
gap.

For a spatial eigenmode \(\phi_\alpha\), define
\(Q_\alpha^\dagger=\sum_{\bm R}\phi_\alpha(\bm R)
q_{\bm R}^{\dagger}\). The exact branch is
\begin{equation}
 |\alpha;M\rangle\propto(S^+)^{M-1}Q_\alpha^\dagger\ketvac,\qquad
 E_{\alpha,M}=\frac J2\lambda_\alpha.
 \label{eq:exact-soft-mode}
\end{equation}
This follows from the one-quartet hopping problem and \([H_{\mathrm{exact}},S^+]=0\). On finite-range periodic lattices \(\lambda_2\sim L^{-2}\), so the parent is rigorously gapless with \(z=2\). In a symmetry-broken thermodynamic description, this branch is the type-B Nambu--Goldstone mode of the pseudospin ferromagnet. Its real-time excitation is a quadratic magnon; diffusion describes only the auxiliary stochastic or imaginary-time dynamics. Write an allowed electron number uniquely as \(N_e=4M+s\), where \(s\in\{0,1,2,3\}\). Then
\begin{equation}
 E_0(4M+s)=
 \begin{cases}0,&s=0,\\U,&s=1,2,3.\end{cases}
 \label{eq:charged-sector-edges}
\end{equation}
One partially occupied cell attains each nonzero edge. These exact thresholds use the parent penalty \(U\), distinct from the microscopic \(U_0\).

\textbf{Charge normalization and density curvature.---}
For magnetic flux \(\Phi\) threaded through a periodic direction, let
\(\theta_e=e\Phi/\hbar\) be the electron twist, where \(e>0\) is the
elementary charge magnitude. The quartet twist is
\(\Theta=4\theta_e\), and the electron-twist curvature is
\(16\rho_4\). This identifies the local charge-\(4e\) flux response.
With electron hopping restored, the full finite-size flux spectrum is
needed to establish exact periodicity rather than infer it from a local
curvature.

All quartet sectors have zero energy, so for \(1\leq M\leq N_s-1\),
\(E_0(4M+4)+E_0(4M-4)-2E_0(4M)=0\).
Equation~\eqref{eq:dicke-density-curvature} fixes the coefficient
linear in \(\delta_{\mathrm{ep}}\) in the thermodynamic density
curvature. The vanishing curvature is a consequence of this parent's
isotropic-ferromagnet structure, not a generic property of RK states.

\section*{S2. Higher-order microscopic coefficients}
\label{app:beyond-leading}

\textbf{Leading projection and cubic structure.---}
In the anti-aligned basis \(\{|4,0\rangle,|0,4\rangle\}\), the
second-order bond contributions are
\begin{equation}
 H_K^{(2)}=-\frac{K^2}{U_0}
 \begin{pmatrix}1&1\\1&1\end{pmatrix},\qquad
 H_t^{(2)}=-\frac{8t^2}{3U_0}
 \begin{pmatrix}1&0\\0&1\end{pmatrix}.
 \label{eq:second-order-bonds}
\end{equation}
Two intermediate \((2,2)\) states of energy \(2U_0\) produce
\(H_K^{(2)}\), while four \((3,1)\) or \((1,3)\) states of energy
\(3U_0/2\) produce \(H_t^{(2)}\). The aligned states
\(|0,0\rangle\) and \(|4,4\rangle\) receive no hopping correction.
Every \(K^2t\) sequence changes endpoint cell parity and cannot return
to the range of \(P^{(0,4)}\). Restoring every cell occupation modulo
four confines each surviving \(Kt^2\) sequence to one bond, while pure
\(K^3\) and \(t^3\) sequences require triangles and vanish on a
bipartite graph. This gives Eq.~\eqref{eq:kt2-correction}.

\textbf{Two-cell coefficients.---}
In the same convention, the isolated-bond corrections beyond the cubic
terms in Eq.~\eqref{eq:kt2-correction} are
\begin{align}
 \delta J_\perp^{(4)}
 &=\frac{-2K^4+\tfrac{32}{3}t^4}{U_0^3}
 -\frac{4K^2V_{\mathrm{nn}}}{U_0^2},\nonumber\\
 \delta J_z^{(4)}
 &=\frac{-2K^4+\tfrac{32}{27}t^4+\tfrac{64}{27}K^2t^2}{U_0^3}
 -\frac{(4K^2+\tfrac{32}{3}t^2)V_{\mathrm{nn}}}{U_0^2}.
 \label{eq:fourth-order}
\end{align}
The \(V_{\mathrm{nn}}\) terms count as fourth order since \(V_{\mathrm{nn}}=O(\lambda^2U_0)\). The \(t^4\) transverse coefficient gives \(t_b^{(t)}=16t^4/(3U_0^3)\); charge-transfer parity forbids odd orders in the pure-electron bond problem, whose next correction is \(O(t^6/U_0^5)\).

\textbf{Bond-truncated easy-plane window.---}
For \(K>0\), Eq.~\eqref{eq:kt2-correction} increases both couplings, more strongly the transverse one. Solving \(J_z(V_{\mathrm{nn}})=\pm J_\perp(V_{\mathrm{nn}})\) with the density-dependent fourth-order terms retained yields the bond-truncated interval width
\begin{equation}
 W_{\mathrm{ep}}^{\mathrm{bond}}
 =\frac{K^2}{4U_0}+\frac{10Kt^2}{9U_0^2}
 +\frac{\tfrac43t^4+\tfrac13K^2t^2-\tfrac18K^4}{U_0^3}.
 \label{eq:window-width}
\end{equation}
The relative transverse correction is \(40t^2/(9KU_0)\), which requires
\(t^2\ll|K|U_0\). The \(V_{\mathrm{nn}}\)-dependent fourth-order terms
shift both edge equations and generate the additional \(K^4\) and
\(K^2t^2\) contributions in Eq.~\eqref{eq:window-width}.

\textbf{Connected-cluster corrections.---}
At \(V_{\mathrm{nn}}=0\), an open three-cell chain additionally generates a next-nearest-neighbor quartet-transfer matrix element \(K^4/(4U_0^3)\). The next-nearest-neighbor density terms have coefficients \(K^4/(2U_0^3)\), \(448t^4/(135U_0^3)\), and \(8K^2t^2/(3U_0^3)\). Density-dependent hopping and the irreducible three-body diagonal term vanish on that cluster. On triangular loops, pure \(K^3\) and \(t^3\) terms occur; the bond-only cubic reduction in the Letter is restricted to bipartite graphs.

Together, the exact two-cell coefficients and open-chain terms define
the present controlled truncation. For the isolated two-cell block, the
first omitted terms are \(O(\lambda^5U_0)\). On an extended lattice,
connected four-cell paths, star clusters (one central cell joined to
three neighbors), and plaquettes can enter already at
\(O(\lambda^4U_0)\) and remain to be computed. The Letter therefore
retains the full local remainder
\(\|\delta H_{\mathrm{rem}}\|_{\mathrm{loc}}=O(\lambda^4U_0)\).

\section*{S3. Stability and dimensionality}
\label{app:infrared-bounds}

\textbf{Density curvature and the \(z=2\) crossover.---}
In a uniform pseudospin coherent state the energy per bond is
\begin{equation}
 \varepsilon_{\mathrm b}=
 -J_\perp\nu_4(1-\nu_4)+J_z(\nu_4-\tfrac12)^2
 =-\frac{J_\perp}{4}+\delta_{\mathrm{ep}}(\nu_4-\tfrac12)^2.
 \label{eq:uniform-energy-curvature}
\end{equation}
Equation~\eqref{eq:uniform-energy-curvature} shows that negative detuning
favors the polarized endpoints and phase separation at fixed
intermediate filling, while positive detuning supplies positive density
curvature. Equation~\eqref{eq:dicke-density-curvature} makes the
linear-in-detuning coefficient exact at the parent, with
\(O(N_s^{-1})\) corrections. The parent therefore anchors the
flat-equation-of-state boundary, and positive detuning selects its
homogeneous side in the leading XXZ model.

The number-phase Berry term is
\(\ii n_4\partial_\tau\phi=\ii\nu_4\partial_\tau\phi+\ii\delta n_4\partial_\tau\phi\).
Only for smooth, topologically trivial configurations is its background part a total derivative. The density-gradient coefficient at the parent is \(c_n=J_\perp/[4\nu_4(1-\nu_4)]\). Equation~\eqref{eq:hydrodynamic-dispersion} then gives \(\omega=J_\perp|\bm k|^2/2\) at zero curvature, matching the exact branch.

\textbf{\(z=1\) phase theory and compact defects.---}
Equation~\eqref{eq:detuning-rg} gives the relevant \(z=2\) flow and
the crossover length for positive curvature. Integrating out the
smooth density field gives
\begin{align}
 \mathcal S_\phi&=\frac12\int d\tau\,d^dx
 [\kappa_4(\partial_\tau\phi)^2+\rho_4(\bm\nabla\phi)^2+\cdots],\nonumber\\
 v_4^2&=\rho_4\kappa_4^{-1}.
 \label{eq:phase-only-action}
\end{align}
Charge \(U(1)\) permits only derivative interactions. After rescaling
imaginary time by \(v_4\), let \(\partial_\mu\) denote derivatives over
the \(d+1\) isotropic Euclidean spacetime coordinates. At the \(z=1\)
Gaussian theory, \([\partial_\mu\phi]=(d+1)/2\), so an analytic vertex
containing \(n>2\) such factors has RG eigenvalue
\begin{equation}
 y_n=(d+1)(1-n/2)<0.
 \label{eq:goldstone-power-counting}
\end{equation}
Equation~\eqref{eq:goldstone-power-counting} is negative for every
\(n>2\), and additional derivatives make a vertex more irrelevant.
Compact vortices and
phase slips are not included in this power counting; they retain
filling-dependent Berry phases. Equation~\eqref{eq:isotropized-stiffness}
gives their control parameter after rescaling time. The compact theory
therefore enters the ordered large-stiffness basin for \(d\geq2\);
defect fugacities are exponentially small in this parameter. This
compact-defect control completes the ODLRO argument for the microscopic
wedge.

\textbf{Controlled microscopic wedge.---}
On a bipartite hopping graph, the particle--hole transformation
\(c_{\bm R,m}\mapsto\epsilon_{\bm R}c_{\bm R,m}^{\dagger}\), with
\(\epsilon_{\bm R}\epsilon_{\bm R'}=-1\) on each bond, leaves
Eq.~\eqref{eq:microscopic} invariant and reverses
\(\hat n_{\bm R}-2\) and \(T_{\bm R}^z\). At zero chemical potential,
it forbids a quartet-density field but does not alone pin the
unrestricted ground state to half filling, particularly at the
cross-sector-degenerate parent.

After the cubic correction is absorbed, choose a positive detuning
satisfying
\begin{equation}
 \lambda^4U_0\ll V_{\mathrm{nn}}-V_{\mathrm{nn}}^{*,(3)}
 \ll\lambda^2U_0
 \label{eq:microscopic-detuning-wedge}
\end{equation}
The hierarchy in Eq.~\eqref{eq:microscopic-detuning-wedge} realizes
Eq.~\eqref{eq:controlled-stability-wedge}, up to fixed coordination
factors, together with convergence of the dressed-operator expansion.
For \(k\lesssim k_\times\), positive curvature provides the leading
restoring force. For \(k_\times\lesssim k\ll1\), the parent
density-gradient term \(c_nk^2=O(J_\perp k^2)\) dominates the local
remainder; at lattice momenta bounded away from zero, the positive
exact-parent scale \(O(J_\perp)\) does the same. These overlapping
regimes exclude a finite-wave-vector density instability for
sufficiently small \(\lambda\).
Within the boundary layer where the remainder competes with curvature,
the renormalized curvature and microscopic transition remain
unresolved. Away from the near-parent domain, finite-wave-vector density
order and charge-\(2e\) gap closure require separate analysis.

\textbf{One-dimensional limit.---}
For the leading one-dimensional nearest-neighbor chain at zero
pseudospin magnetization, \(-1<\Delta<1\) gives a quartet Luttinger
liquid \cite{Luther1975,Haldane1980,Giamarchi2003}. Write
\(\Delta=\cos\gamma\), with \(0<\gamma<\pi\), and let \(K_L\) denote the
quartet Luttinger parameter. Then
\begin{equation}
 \langle q_0^\dagger q_{\bm R}\rangle\sim|\bm R|^{-1/(2K_L)},
 \qquad \frac1{2K_L}=1-\frac{\gamma}{\pi}.
 \label{eq:luttinger-exponent}
\end{equation}
Equation~\eqref{eq:luttinger-exponent} shows that quartet correlations
dominate density correlations at wave vector \(2k_F\) throughout the
strict interval, where \(k_F\) is the Jordan--Wigner-fermion Fermi
momentum. Pair correlations remain short ranged in the convergent
microscopic regime. The exponent approaches zero at
\(\Delta\to-1^+\), but the endpoint itself is the isotropic ferromagnet,
not a conformal Luttinger liquid. Higher-order terms or filling away
from half renormalize the exponent. In a stable phase, true quartet
ODLRO and an extensive four-particle Yang eigenvalue require \(d\geq2\).

\section*{S4. Relation to existing constructions}

\textbf{\(\eta\)-clustering and Richardson--Gaudin context.---}
Yoshida and Katsura constructed even-\(N\) fermion
\(\eta\)-clustering states and a parent for their ground-state tower
\cite{Yoshida2022}. On a bipartite lattice, let
\(\epsilon_{\bm R}=\pm1\) on the two sublattices. At \(N=4\), the
transformation
\(q_{\bm R}^{\dagger}\mapsto\epsilon_{\bm R}q_{\bm R}^{\dagger}\)
maps their staggered tower to Eq.~\eqref{eq:dicke-state}, identifying
the quartet Dicke tower as the \(N=4\) member of their construction.
The present extensions are the minimal uniform parent, its
graph-resolved spectrum and phase response, and its connection to the
lower-body mechanism. Yang's \(\eta\)-pairing construction and
pair-hopping ground-state models supply the earlier lineage
\cite{Yang1989,deBoer1995,Montorsi1996}.

In each fixed-\(M\) sector, the same state is the ground state of the
degenerate quartet analogue \(H_{\mathrm{RG}}=-G S^+S^-\), with
coupling \(G>0\) and energy \(-GM(N_s-M+1)\), in the
Richardson--Gaudin family \cite{Richardson1963,Gaudin1976,Dukelsky2004}.
On the complete graph, the projected parent is equivalent to
\(H_{\mathrm{RG}}\) up to an overall scale and a sector-dependent
constant. On generic finite-range graphs, exact solvability instead
follows from frustration freeness and exclusion dynamics. Related
quartet correlations occur in nuclear spin--isospin pairing and
\(\alpha\)-particle condensation
\cite{Dukelsky2006,Lerma2007,Ropke1998,Tohsaki2001}, and in
four-component one-dimensional gases \cite{Wu2005,Lecheminant2005}.

\textbf{Filling constraints.---}
The exact parent's gaplessness is consistent with the Lieb--Schultz--Mattis--Oshikawa--Hastings filling constraints \cite{Lieb1961,Oshikawa2000,Hastings2004}. It conserves every flavor number, each at noninteger filling \(\nu_4\), so a unique, gapped, translation-invariant short-range-entangled state preserving \(U(1)^4\) is excluded for \(0<\nu_4<1\). If only total charge is conserved, the filling obstruction instead requires \(4\nu_4\notin\mathbb Z\). The exact \(z=2\) branch at all intermediate fillings follows from pseudospin symmetry and exclusion dynamics, not from a universal total-charge filling obstruction.

\textbf{Other microscopic routes.---}
Other lower-body models realize complementary direct quartet phases.
Soldini, Fischer, and Neupert established a one-dimensional quartet
liquid beyond the perturbative limit \cite{Soldini2024}; recent quantum
Monte Carlo (QMC) studies find zero-temperature charge-\(4e\) order in
attractive \(SU(4)\) and SSH-interaction models \cite{Wan2026,Shi2026}.
For the \(N\)-flavor SSH model, let \(i,j\) label lattice sites,
\(\alpha=1,\cdots,N\) label flavors, and \(J_{\mathrm{SSH}}>0\) denote
the bond-interaction strength. Define
\begin{equation}
 \begin{aligned}
 B_{ij}&=\sum_{\alpha=1}^N
 (c_{i\alpha}^\dagger c_{j\alpha}+\mathrm{H.c.}),\\
 H_{\mathrm{SSH,int}}&=-\frac{J_{\mathrm{SSH}}}{2N}
 \sum_{\langle ij\rangle}B_{ij}^{\,2}.
 \end{aligned}
 \label{eq:ssh}
\end{equation}
Expanding Eq.~\eqref{eq:ssh} yields pair transfer together with density,
exchange, and correlated hopping. The connection is operator-level:
because the SSH strong-coupling manifold is bond centered rather than
the onsite empty/quartet doublet used here, a quantitative mapping from
\(J_{\mathrm{SSH}}\) to \(K\) and \(U_0\) requires a separate
strong-coupling projection. The reported finite-temperature
two-dimensional quartet phase has quasi-long-range order, distinct from
the zero-temperature ODLRO considered in the Letter.

\end{document}